# Probability Density Function Estimation in OFDM Transmitter and Receiver in Radio Cognitive Networks based on Recurrent Neural Network

*Mahdi Mir\**,
*Department of Electrical Engineering, Ferdowsi University of Mashhad, Mashhad, Iran*

*Abstract*—**The most important problem in telecommunication is bandwidth limitation due to the uncontrolled growth of wireless technology. Deploying dynamic spectrum access techniques is one of the procedures provided for efficient use of bandwidth. In recent years, cognitive radio network introduced as a tool for efficient use of spectrum. These radios are able to use radio resources by recognizing surroundings via sensors and signal operations that means use these resources only when authorized users do not use their spectrum. Secondary users are unauthorized ones that must avoid from interferences with primary users transmission. Secondary users must leave channel due to preventing damages to primary users whenever these users discretion. In this article, spectrum opportunities prediction based on Recurrent Neural Network for bandwidth optimization and reducing the amount of energy by predicting spectrum holes discovery for quality of services optimization proposed in OFDM-based cognitive radio network based on probability density function. The result of the simulation represent acceptable value of SNR and bandwidth optimization in these networks that allows secondary users to taking spectrum and sending data without collision and overlapping with primary users.**

Introduction

\* Corresponding Author.
E-mail address: Mahdimir.ir@gmail.com (Mahdi Mir)

In the last two decades, requesting spectrum had grown due to wireless Services and products. Todays, according to frequency allocation almost all band frequencies allocated. Cognitive radio network suggested as a solution for overcoming Lack of spectrum resources in order to providing radio resources for new wireless applications. Static spectrum allocation strategy caused time slots and geographical in allowed bands spectrum. One of the important points in this networks are how to transit frequency. This thing happens when primary users send their spectrum and due to lower priority of secondary users, they must leave the spectrum and must search the channels for finding empty spectrum. Due to the direct disassociate between primary and secondary users, restricting secondary users to decision based on local spectrum sensing for reasons such as multi-path fading will be greatly reduced performance. Using one sensor for spectrum detection by using recognizing the primary user transmitter may be faced with two types of problem. First, primary transmitter is not in the secondary detection range and second is hidden terminal problem. The rest of this article is as follows: in second part we will describe basic OFDM-based cognitive radio networks concept briefly. In third part, we will examine some recent works. In fourth part, proposed method will be describe. In fifth part, results of simulation represented. In sixth part, conclusion and in seventh part, some ideas are named for future researches.

OFDM-based Cognitive Radio Networks and Methods

Cognitive radio network used as a solution for spectral congestion problem that works with using opportunistic frequency bands that does not occupied by authorized users. Cognitive radio defined according to Federal Communication Commission definition: "a radio or system that sense the operational electromagnetic area and can be adapt radio parameters automatically to optimize system operation such as maximizing throughput and minimizing interference [1].

OFDM-based Cognitive radio network contains primary and secondary network that taking in a geographical place. Primary network contains several authorized broadband. According to researchers in spectrum, the occupancy rate of these authorized bands is relatively low. Secondary network is based on infrastructure that several spectrum sensors such as intelligent phones or tablets and synthesize base station for detecting status of authorized bands cooperate each other.

Intelligent radio defined by using extraction spectral opportunities in both authorized and unauthorized band as solution for optimizing general application of spectrum. Smart radio cycle started with the aim of gaining information from wireless area to selecting the best channel by opportunistic users. These nodes that authorized in specified spectral bands are primary users and they have higher and legal priority in using special part of spectrum. Secondary users are obliged to leave the channel due to primary user's quality of services without damage and they have lower priority of using spectrum without any collision and overlapping with primary users. So, secondary user needs to have spectral sensing ability that can be detect primary user presence. Primary users do not have any worries about behavior of intelligent network and there is no special features for coexistence. Secondary users do not have certificate usually and should not interference with primary user transmission. So, whenever primary user recognized, secondary user must react instantly with changing power of RF, used channel rate and etc. It's because their transmission should not reduce quality of services access to spectral channel and avoid collision between intelligent users. In fact, intelligent radio network provide a solution for spectral allocation between users.

MAC plays an important role in several cognitive radio operation such as spectrum mobility, channel sensing, MAC allocation and spectrum sharing. Spectrum mobility allows the secondary users to leave their channel and access to free band and connect a link again when primary users recognized. Channel sensing is the ability of intelligent users for collecting information about spectrum application and maintain a dynamic view of available channels. Resource allocation is for opportunistic allocation of available channels to intelligent users with their quality of services requests. Spectrum access deals between primary user and heterogeneous secondary for preventing interference.

MAC protocols of OFDM-based cognitive radio network categorized with features like complexity, protocol architecture, cooperate level in network, how to manage signaling and data transmission in communication. MAC protocol have two main category means DAB and DSA.

Spectrum Sensing with Probability Density Function

The main part of OFDM-based cognitive radio network is spectrum sensing that is knowledge about the use of spectrum and available main user in the geographical place and it is an agent that enables intelligent radio to get information and recognizing spectrum opportunities with probability density function. Though, spectral spectrum sensing recognized as traditional to sense spectral content or energy sensing of spectrum radio frequency, but in cognitive radio contains achieving spectral features in multiple dimensions such as time, place, frequency and code. It contains determination type of signal that occupied spectrum such as modulation, wave form, carrier frequency and etc. anyway, these requirements leading to analyze powerful techniques of signal with additional computing complexity.

When two nodes decided to communicate each other, source and destination have to sense it. They select the set of channels for sensing, then estimate the availability of channel, then filtering done and at last, communication link tuning. Reaction and precaution sensing can be used in intelligent network. Sensing period within a range can be done during data transmission to recognizing primary users and preventing collision.

Spectrum protocols categorized in two general categories means narrow band spectrum and bandwidth spectrum. Matched filter, energy detection and detection based on circulating static properties. The term narrow band refers to frequency range that is narrow enough to consider flat answer of channel frequency [1]. Spectrum sensing contains spectral sensing based on energy detection, based on wave form and based on circulating static properties. This article and proposed method is spectral sensing based on energy detection.

The most common spectral sensing is energy detection that recognized as radiometric method, too. In this method, signal in comparison with energy output with a threshold value that depends on noise value will be detected. It's assumed that received signal is as (1).

$$y(n) = s(n) + w(n) \quad (1)$$

According to (1), $s(n)$ is detected signal, $w(n)$ is additive white Gaussian noise (AWGN) and n is samples indices. Decision criteria for energy detection is as (2).

$$M = \sum_{n=0}^{N} |y(n)|^2 \quad (2)$$

According to (2), N is observed vector size. The decision to occupying a band can be done with comprise M criteria with a fixed threshold value $\lambda_E$. The assumption is equivalent to two as (3).

$H_0: y(n) = w(n),$

$$H_1: y(n) = s(n) + w(n). \quad (3)$$

The performance of detection algorithm with two probability: 1) the probability of correct detection or $P_D$, 2) the probability of false detection or $P_F$. In this case, $P_D$ on intended frequency is a signal and formulated as (4).

$$P_D = \Pr(M > \lambda_E | H_1) \qquad (4)$$

According to (4), $P_F$ is a false detection and recognize busy frequency and if it is not busy, can be shown as (5).

$$P_F = \Pr(M > \lambda_E | H_0) \qquad (5)$$

The threshold value can be select optimal between $P_D$ and $P_F$. For more detailed, $P_D$ and $P_F$ can be shown as (6).

$$P_D = \Pr(Y > \lambda_E | H_1) = Q_u(\sqrt{2r}, \sqrt{\lambda})$$

$$P_F = \Pr(Y > \lambda_E | H_0) = \frac{\Gamma_i(u, \frac{\lambda}{2})}{\Gamma(u)} \qquad (6)$$

According to (6), r is SNR and $\lambda$ is threshold value. $\Gamma_i(u, \frac{\lambda}{2})$ is incomplete gamma function and $\Gamma(u)$ is complete gamma function. $Q_u$ is Q macron function.

Energy detection can be done in time or frequency domain. For signal sensing in special frequency band in time domain, a band pass filter apply to signal and measure the output signal samples power. For measuring signal power in frequency domain, FFT take and signal energy measurement in whole part of signal at target frequency area.

Challenges of OFDM-based Radio Cognitive Networks

One of the main challenges of OFDM-based cognitive radio network is recognizing holes place or spectral opportunities in a place and when it take place. Spectrum sensing is the key technology for opportunities recognizing. Primary user does not have any necessarily for sharing and changing their operational parameters for spectrum sharing with cognitive radio. Therefore, cognitive radio should detect spectral opportunities without help of primary user.

One of the criteria of quality of services is secondary user throughput that can determine system design frameworks. Spectral detection needs significant energy consumption. This amount of consumption can be reduce. By providing inactive channels. The ability to allow secondary user for spectrum sensing will have. By using this method, optimizing spectrum can be done. It's necessary to reduce predicting error probability that it will be done at this research by Recurrent Neural Network method.

Statistical hypothesis testing in a method in statistical science that examining distribution parameters in statistical population. In this method, zero assumption or primary assumption will be examined that proportional to hypothesized study select as alternative hypothesis to each correctly relative to each other tested.

Statistical hypothesis testing decide between true hypothesis and its complement. In other words, to test the $H_0: \theta \epsilon \ominus_0$ hypothesis against its complement $H_1: \theta \epsilon \ominus_0^c$ and likelihood ratio described as (7).

$$\lambda(x) = \frac{\text{SUP}_{\theta \epsilon \ominus_0} \delta(\theta|x)}{\text{SUP}_{\theta \epsilon \ominus_0^c} \delta(\theta|x)} \qquad (7)$$

According to equation (7), $\delta(\theta|x)$ is likelihood ratio of data. A likelihood ratio test in a test that $H_0$ assumption denies $\lambda (x \leq c)$. The Neyman Pearson method defined that likelihood ratio test for constantly simple hypothesis test $H_0: \theta \epsilon \theta_0$ is the most powerful hypothesis test.

Pearson's Chi-Squared test is a statistical test of nonparametric due to some size evaluation of nominal variables that defined as (8).

$$X^2 = \sum_{t=1}^{m} \frac{(O_t - E_t)^2}{E_t} \qquad (8)$$

According to (8), O is observed abundance and E is expected abundance. This test is without distribution and expected abundance should not be zero. Total categories that their observed value is less than 5, should not be more than 20 percentage of all categories. This test is only solution of homogeneity test about nominal scale variables with more than two categories. So, it has more frequent usage in comparison of other statistical test. This test is sensitive to sample size.

Literature Review

Many articles have been discussed to ratio cognitive scenarios until now. At these articles, types of networks like MANET and cellular network considered as secondary network. In [2] decentralized cognitive MAC protocols suggested that allow secondary users to search independently and without usage of central coordinator or specified telecommunication channel for finding opportunistic spectral. In this article, an analytically framework for random access to developed spectrum that is based on Markov Chains Decision Making. In [3 and 4] routing protocols and channel allocation in MANET cognitive radio surveyed.

In [5] using Bayesian learning for predicting an available channel or not available proposed that learning is considered geometric distribution for simple channel usage pattern. Sensing selected channels for finding the best channel for sending is depends on number of considered channels. Intuitively, a smaller number of selected channels caused to reducing total sensing time, but an empty channel for sending between them is so hard. So, it seems that there is comprise between total sensing time in a range and an empty channel probability between selected channels. Therefore, it is a challenge to find optimized value for the number of selected channels in order to minimizing total time of sensing system. In a research, efforts to determine selected optimized channels. First, a subset of candidate channels can be used as secondary network channel defined and then described when take a place at candidate channels. This method called channel awareness [5, 6, and 7].

Decision making scheme based on sensing and probability to distribute secondary user data between several channels proposed [8, 9, and 10]. At these articles, the optimized number of selected spectrums and optimized value of channel selection probability based on probability proposed. The purpose of these two schemes is minimizing total time of secondary user system. Suggested algorithms of these works are not to find optimal sequence sensor and designed for only one secondary user at time to use spectrum.

In [12, 13, 14, 15, and 16] neural networks proposed for using in cognitive radio network. Secondary user sorted frequency spectrums in the specified form. Then optimization problem formulate in order to maximizing secondary user throughput. Studying consumed energy for sensing process show ineffectiveness of throughput maximizing. To fix this problem, a throughput maximizing problem formulated based on energy. This optimization value due to high complexity have modeling limitations. It means that to reach to optimized answer based on theorem needs heavy simplifier assumption. For these reason, artificial neural networks used as powerful tool to modeling and optimizing complex systems and time variant. Two kinds of neural networks means KC and MFF used. Their features discussed in complex systems modeling and solving optimization problems and time variant. Based on these proposed neural networks, a structural method proposed to determine the optimal sensing time.

In [17] network flow approach proposed for network selection to examine secondary user in cognitive radio network. Network selection operations a problem that there are very little research about them. In this approach, a network flow framework proposed to select network and users. Results show that this approach can allocate again secondary users and channel at the same selected network that maximize quality of services to primary and secondary users.

In [18] a layered and opportunistic multi-channel MAC protocol proposed to spectrum sensing in physical layer with packet sending scheduling in network. In this method, each secondary user equipped with more than two transmitter and receiver that one of them is for periodic sensing in cognitive radio network and other one is for recognizing unused channels. This method works dynamically. Results show better understanding of unused channel for spectrum sensing by equipment of secondary users.

In [19] performance analysis of cognitive radio network proposed by considering users quality of services and channel certainty. Channel estimation and quality of services proposed to primary and secondary users by using MMSE method and result show that the maximum channel capacity is used.

In [20] operations and rules of primary and secondary users in cognitive radio network proposed to prevent overlapping. They developed two user operation to transmit data from primary user with maximizing area to make some rules between users to measure spectrum in high security at an online dynamic area.

In [21] analyzed existing model of transmitting spectrum decision making system proposed in a cognitive radio network that can optimize primary user performance in sending and decision by cutting the connection of secondary user with unrecognizing area. The main results of this research is reducing delay and sending data time from primary user.

In [22] control power for primary and secondary user optimized in cognitive radio network that caused to make a new cost function for power transmit and total of that for optimizing Nash equilibrium point. The results show that proposed algorithm with new cost function achieved for maximizing access number of secondary user to unused spectrum and developed primary and secondary user application in cognitive radio network.

In [23] supply the quality of services for heterogeneous services proposed in cognitive radio network. In this approach, a spectrum allocation framework proposed for supply the quality of services for secondary user access in heterogeneous services and real-time process for spectrum sensing, decision spectrum access, channel allocation and control admission in distributed operation of cooperative cognitive radio network. The results show that proposed approach is helpful to optimizing quality of services in spectral resources allocation.

In [24] a method proposed to spectrum selection for quality of services satisfaction in cognitive radio network that results show that channel selection scheme caused reducing total transmit time of secondary user. Total transmit time caused increase received rate from secondary user and spectrum sensing based on channel selection scheme have high received performance and rate by secondary user in asleep time of primary user or even in real-time process. This proposed channel selection scheme can help secondary user data transfer even in multi-channel states.

In [25] leading quality of services proposed in a route with channel selection for heterogeneous cognitive radio network that secondary user wants to exploit resources from primary user. The main idea of this network is channel selection of secondary user for finding spectral opportunities with lowest cost that caused optimized quality of services to secondary users.

In [26] secondary user cooperation proposed in cognitive radio network for precision balance of spectrum sensing and exploiting. The proposed method is the combination of sequential methods, semi-parallel, simultaneous and asynchronous in cooperative spectrum sensing of cognitive radio network users that caused to achieve new rate of transmission and spectrum sensing.

In [27] supporting optimized quality of services proposed for secondary users in cognitive radio network. Spectrum stability and capacity usage of that by primary and secondary user is the main idea to present optimal services to secondary users with minimum bandwidth. The results show that quality of services optimized for secondary user and overlapping in operations and reducing spectrum collision.

The classic test of probability [28 and 29], energy detection [30 and 31], detection based on adaptive filters such as RMS, LMS and NLMS [32], detection based on circulating static features [33, 34, and 35] and many newfound methods [36] are the examples of spectrum detection methods have been studied until now.

Proposed Approach

Neural networks are nonlinear parametric models that make a mapping function in input and output data. Input data of Recurrent Neural Network use a TV band. TV band can be TV broadcasting system with multi-antenna. The proposed method use binary series if input and output. Binary series defined as (9).

$$X_1^T = \{x_1, x_2, \ldots\ldots, x_t, \ldots\ldots, x_T\} \quad (9)$$

According to (9), for each channel with sensing status, a channel will be produce in each gap and in T time duration. Channel status in each gap is busy or not busy that will be show with 1 and $-1$ binary signs. Recurrent Neural Network predictor learn by using binary series and based on this learning, cannel status in next gap will be predict by gap status records. Predictor allocate for each channel in a multi-channel system.

Energy detection is a suitable method for channel spectrum sensing, but under the term of condition it is possible. Channel spectrum sensing considered as an assumption between $n_i$ noise and $s_i$ signal. When bandwidth B, complex NB in N seconds in time takes long. These assumption described as (10).

$H_0: y_i = n_i \quad , \quad i = 1,2, \ldots, NB$

$$H_0: y_i = n_i + s_i \quad , \quad i = 1,2, \ldots, NB \quad (10)$$

It's assumed that Gaussian distribution $s_i$ is like a complex circular symmetric with mean zero and $\sigma_s^2$ variance. Another assumption is independent sample signal $\{s_i\}$ recognized as distributors. With these assumption to detect above theory, Neyman Pearson used as (11).

$$Y = \frac{1}{NB}\sum_{i=1}^{NB} |y_i|^2 \underset{<}{\overset{>}{H_0^1}} \lambda \quad (11)$$

According to (11), $\lambda$ is detection threshold. Y's observation is based on Pearson's Chi-Squared with two degree freedom 2NB. We can create a probability of detected sample or false one as (12) and (13).

$$P_f = \Pr\{Y > \lambda | H_0\} = 1 - P\left(\frac{NB\lambda}{\sigma_s^2}, NB\right) \quad (12)$$

$$P_d = \Pr\{Y > \lambda | H_1\} = 1 - P\left(\frac{NB\lambda}{\sigma_n^2 + \sigma_s^2}, NB\right) \quad (13)$$

According to (12) and (13), $P(x,y)$ is to tuning gamma function as $P(x,y) = \frac{\lambda(x,y)}{\delta(y)}$ that $\lambda(x,y)$ is the lower bound of gamma complement and $\delta(y)$ is gamma function. Due to exploiting from each possible channel with spectral opportunities, secondary user cooperates for recognizing unidentified channels by using certain spectrum sensing policy. Each user determine spectral opportunities and then considered set of L channels sequentially based on channel dynamic one dimensional with spectrum sensing time $t_s$. Then each of them report their observation about channel in cognitive radio network. The optimized spectrum sensing channel for each user is as (14) by considering primary user intervention and take spectrum from user.

$$L = \min\left(\left\lceil \frac{N}{M_s} \right\rceil, \left\lceil \frac{T - T_c}{2t_s} \right\rceil\right) \quad (14)$$

According to (14), $\lceil x \rceil$ is the nearest integer number greater or equal to x and $M_s$ is the number of users that try to sensing spectrum for an opportunity. N is the number of channels. T is the time range and $T_c$ is time duration of MAC message controlling that calculate as (15).

$$T_c = T_{B1} + T_{B2} + NT_{ms} + 5T_{SIFS} \quad (15)$$

According to (15), $T_{B1}$ and $T_{B2}$ are time duration of $B_1$ and $B_2$ nodes. $NT_{ms}$ is time duration of report for N channel in OFDM-basedcognitive radio network. $T_{SIFS}$ is SIFS unit for delay propagation and time for spectrum sensing of each channel that calculate as (16).

$$t_s = \left(\frac{\sqrt{2\lambda + 1}Q^{-1}(P_d) - Q^{-1}(P_f)}{\lambda\sqrt{B}}\right)^2 \quad (16)$$

According to (16), $Q(x) = \frac{1}{\sqrt{2\pi}} \int_x^\infty \exp\left(\frac{-t^2}{2}\right) dt$ and $P_d$ and $P_f$ are detection probability and false detection threshold that defined by primary and secondary user. B is channel bandwidth and $\lambda$ is SNR sensitivity value to detected spectrum. N channel detected slightly based on number of users try to sensing spectrum to have an opportunity. It must be determined that the number of users that do not have access to channel for channel resources allocation for optimizing quality of services. A channel considered that primary user have access certificate to that. A channel can be active or inactive at any moment. It is assumed that active or inactive time is exponential distribution. Secondary users equipped as opportunist user with a cognitive radio and some sensors. $S(n)$ used as channel status in n slot. If channel is in inactive status, $S(n) = 1$ and if channel is in active or busy status, $S(n) = 0$. So, we have (17).

$P(S(n+1) = 0 | S(n) = 1) = q$

$$P(S(n+1) = 0 | S(n) = 0) = p \quad (17)$$

According to (17), $P(q)$ is transition probability of active (inactive) to inactive (active) state. $P|(p+q)$ is channel stability state probability for inactive status. Logical assumption is $1 - p -$

$q > 0$ that implicate with adjacent slot that have the most similarity to channel. Spectrum sensors detect primary user signals by channel sensing. Spectrum sensors report channel information of OFDM-based cognitive radio network that try to decide channel access. These decision is from A operational area that is described as (18).

$$A = \{D:0\text{(Transmitting)}, 1\text{(Sensing)}, 2\text{(Sleeping)}\} \quad (18)$$

According to (18), 0 is for transition, 1 is for sensing and 2 is for active or inactive state. For example in initial state, $n, D_n = 2$ means transmission started from secondary user. Decision based on information apply from different states of channel. In n slot that is $X_n$, the probability of channel in n slot that is inactive estimated. $X_n$ value from $\{\tau: 0 \leq \tau \leq 1\}$ is countable set in 0 and 1 range. $X_n$ updated based on rules. When $D_n = 0$ or 2 we have (19).

$$X_{n+1} = X_n(1-q) + (1-X_n)p \quad (19)$$

And when $D_n = 1$ we have (20).

$$X_n + 1 = \begin{cases} 1, \text{w. p.} X_n(1-q) + (1-X_n)p; \\ 0, \text{w. p.} 1 - X_n(1-q) - (1-X_n)p \end{cases} \quad (20)$$

That in $1, \text{w. p.} X_n(1-q) + (1-X_n)p$ time when channel is in active or inactive state, secondary users updated by Markov transition. In $0, \text{w. p.} 1 - X_n(1-q) - (1-X_n)p$ time, it is an assumption that the best result of channel sensing achieved that is 1 for inactive and 0 is active. In $X_n(1-q) + (1-X_n)p$, a $n+1$ slot probability sensed as inactive. It is assumed that one packet sent from slot. In n slot if $S(n) = 1$ means packet sent successfully. In this case, secondary user have a certain operational power that can be transmit by defining $R_t$. If $S(n) = 0$, the collision will be happen when secondary user packet hit with primary user's packet. In this case, secondary user penalties by a $C_c$ cost. Whenever cost of penalty is greater, safety of primary user increased, but reduced transmission opportunities for secondary users. If $X_n$ information state at the beginning of slot equal to $\tau$, total reward for transmission is $\tau R_t - (1-\tau)C_c$. It should be noted that if channel probability average is $\frac{p}{p+q}$, it will measured from database. Secondary user often have no information about channel status before channel access. If $\frac{p}{p+q}R_t - \frac{1}{p+q}C_c > 0$ secondary user always have transition and reception without channel sensing, but it will challenge the security of primary user. So, it is reasonable to suppose $\frac{C_c}{R_t+C_c} \geq \frac{p}{p+q}$ take happen. If $D_n = 1$ secondary user try to sensing channel in $n$ slot. In this case, subsidiary and main approach created that is not a topic in this research and the best status will be considered. Channel sensing allow secondary user that have exact information from channel and this is vital work for sending or not sending packet by considering channel status. In fact, an opportunistic method created to estimate channel status by secondary user and if conditions were appropriate and primary user have not channel, try to send data over network. Another important tip is that channel sensing needs cost. If secondary user have limited energy, channel sensing try to polluting resources for both active and inactive sensing and a lot of energy waste. If at this moment that channel sensing is trying to polluting resources, secondary user try to send packet, collision and overlapping with primary user happen. At this research for simplicity it is assume that sensing cost is fixed that defined by $C_s$.

At this research, inactive status of secondary user considered that in this state there is no reward belongs to secondary user. To briefing explained topics for taking spectral opportunity, described achieving by secondary user at n slot depends on $X_n$ information state and $D_n$ performance that can be written as (21).

$$G(X_n, D_n) = \begin{cases} \tau(R_t + C_c) - C_c, & X_n = \tau, D_n = 0; \\ -C_s, & D_n = 1; \\ 0, & D_n = 2; \end{cases} \quad (21)$$

If τ was great, secondary user must transmit data and if τ was small, secondary user should sleep to find another opportunity for storing sensing costs.

Now a criteria needs for showing useful signal against noisy signal in cognitive radio system. The value less than 12 dB show the serious problem in channels. The value more than 20 dB is satisfying and higher than 30 dB is so suitable. Actually, this index is better and show more useful signal. SNR defined as signal to noise ratio power and calculated as (22).

$$\text{SNR} = \frac{P_{signal}}{P_{noise}} \quad (22)$$

According to (22), P is average value of signal power. Due to the most signals have dynamic range, they described as dB logarithmic that will be as (23) for power signal and (24) for noisy signal.

$$P_{signal,dB} = 10\log_{10}(P_{signal}) \quad (23)$$

$$P_{noise,dB} = 10\log_{10}(P_{noise}) \quad (24)$$

The statistical estimate RMSE used in order to comprise predicted errors by a dataset in OFDM-based cognitive radio network. This method have no application for comprise several dataset. Individual difference are called remaining and RMSE used for storing them in one number. In RMSE, a statistical estimator θ with respect predicted parameters defined as the square root of the mean square errors as calculated in (25).

$$\text{RMSE}(\theta) = \sqrt{\text{MSE}(\theta)} = \sqrt{E((\theta - \theta)^2)} \quad (25)$$

In RMSE, root square is variance and recognized as standard error. RMSE normalization by observed value range calculated as (26).

$$\text{NRMSE} = \frac{\text{RMSE}}{X_{max} - X_{min}} \quad (26)$$

When minimizing value show less variance erosion, usually defined as percentages. Another criteria for evaluation is PSNR that is peak signal to noise ratio and calculated as (27) in terms of dB.

$$\text{PSNR} = 10.\log_{10}\left(\frac{MAX_I^2}{MSE}\right) \quad (27)$$

That $MAX_I^2$ is the great possible signal.

Now that spectral opportunity type determination obtained for secondary user, predicting by Recurrent Neural Network is simple work to optimize quality of services to these users. Neuro-fuzzy model extended to obtain more accuracy result. The output of Recurrent Neural Network model depends on used parameters for learning. Recurrent Neural Network will learn based on considered parameters and will use after learning for prediction. Due to these descriptions, the inputs of Recurrent Neural Network is channel capacity, qualitative efficiency at scanned channel, and distance between primary user base station and secondary user base station. The output of Recurrent Neural Network is channel status. Based on relative input information with channel to Recurrent Neural Network, channel status is $1$ or $-1$ that $1$ represent occupancy of channel and $-1$ represent inactiveness of channel.

Simualtion and Results

MATLAB environment as simulation platform and ANFIS toolbox at command window used. The target is spectral opportunity prediction due to quality of services optimization for secondary user in a OFDM-based cognitive radio network. Data are TV data. The threshold value defined as 0.181. Initial weight is 0.5 and learning iteration of Recurrent Neural Network is 400. Fig. 1 showed the normalized throughput ratio to average secondary user density per $km^2$.

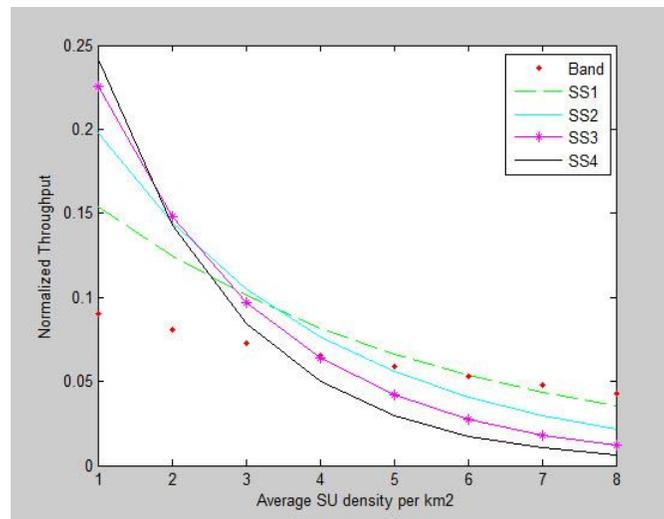

Fig. 1. Normalized throughput ratio to average secondary user density per $km^2$.

In Fig. 1, SS means spectrum sensing for users on main band. Then training and validation based on RMSE in Recurrent Neural Network done that represented in Fig. 2.

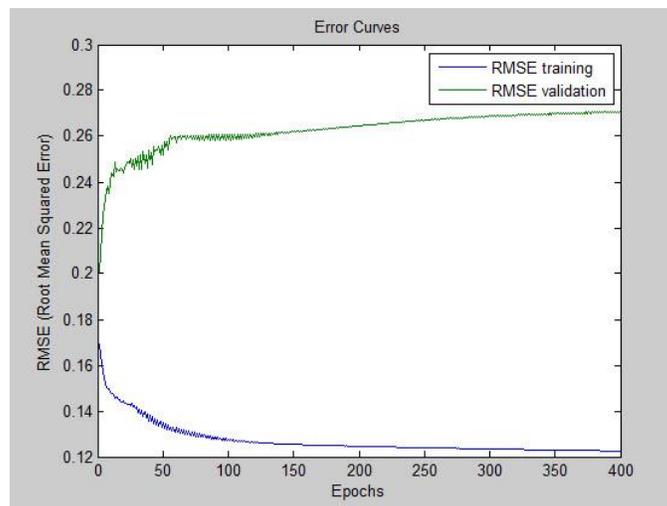

Fig. 2. Training and validation based on RMSE in Recurrent Neural Network

The obtained results for MSE, RMSE, PSNR and SNR represented in Table I.

TABLE I. Obtained Results of MSE, RMSE, PSNR and SNR

| RMSE training | 0.2143 |
|---|---|
| RMSE validation and test | 0.2116 |
| MSE | 0.46 |
| PSNR | 21.4549 |
| SNR | 27/4231 |

Result of SNR based on recent descriptions that if it is less than 12 dB, serious problem in channels, more than 20 dB is satisfying and more than 30 dB is suitable, show a good performance in comparison other methods that examined. In fact, the more of this criteria means better performance. The obtained result of SNR in this approach is 27.4231 dB. With more training, it will be changed between 27 dB to 29 dB. It is clear that channel status detection for providing quality of services to secondary user in OFDM-based cognitive radio network based on predicted spectral opportunities have good and acceptable value.

The proposed framework for managing dynamic spectrum have four process such as spectrum sensing, spectrum decision making, spectrum sharing and spectrum mobility. Spectrum sensing is the ability of determining unused spectrum and it should share the spectrum without any collision with other users. Spectrum decision making is for providing user requirements to select the best existing spectrum. Spectrum sharing is for providing fairy schedule between users to use spectrum and channel. Spectrum mobility is for integration maintaining of user requirements in transmission.

Conclusion

In this article, Recurrent Neural Network method used for bandwidth optimization for better quality of services and energy reduction by predicting spectral opportunities and spectrum holes discovery in OFDM-based cognitive radio network. Based on proposed approach, it can be seen predicting spectrum opportunities in cognitive radio channel with sensing secondary user can optimize quality of services and obtain the acceptance value of SNR in dB. With the help of this approach we will enable to increase bandwidth by determining channel from sensing relation by secondary user due to obtaining an opportunity to send data in channel and decrease the energy consumption in inactive status. So, the proposed method makes it possible to predict channel status to use it optimal.